\documentclass[12pt]{article}
\def\ii{\'{\i}}

\def\be{\begin{equation}}
\def\ee{\end{equation}}
\def\ba{\begin{eqnarray}}
\def\ea{\end{eqnarray}}

\def\fr{\frac}
\def\dl{\partial}
\def\slash{\not\!}

\def\iv{\frac{1}}
\def\ra{\rightarrow}
\newcommand{\bx}
  {\hbox to.77778em{\hfil\vrule\vbox to.675em
  {\hrule width.6em\vfil\hrule}\vrule\hfil}}

\def\ni{\noindent}

\def\vs{\vspace{0.5cm}}

\begin{document}
\thispagestyle{empty}
{\tiny flucmet.tex 20010331}
\vs

\centerline{\huge \bf Fluctuations of the Metric Tensor}
\vs

\centerline{\huge \bf and Fermion Propagators \Large
\footnote{Partially supported by ESO, Praxis, Sapiens, CERN and FCT, grant numbers  
PESO /P/PRO/15127/1999, PRAXIS/P/FIS/12247/1998, POCTI/1999/FIS/35304 and CERN/P/FIS/40119 /2000}}
\vs

\vs

\centerline{Alex H. Blin}
\vs

\centerline{Centro de F\ii sica Te\'orica, Universidade de Coimbra, P-3004-516 
Coimbra, Portugal}

\centerline{\texttt{alex@teor.fis.uc.pt}}
\vs

\vs

\ni {\sl Abstract} - Conformal fluctuations of the metric tensor at the Planck scale are considered. They give rise to a lower bound of the proper length. This leads to finite expressions for quantities related to propagators without the need of renormalization or regularization. Quantities like the current quark mass or the effective strong coupling constant have to be reinterpreted.
\newpage

\section{Introduction}
\vs

The concept of a particle in Quantum Field Theory is point-like. This fact leads to the known divergent expressions, which are usually tackled by renormalization or regularization techniques. Quantum Gravity, however, introduces a coarse graining of space-time at the Planck scale. The idea that quantum fluctuations of the metric can have the effect of a regulator is not new, see for instance \cite{deser} and references therein, and \cite{dewitt,isham}.
\vs

The coarse graining of space-time indicates that the metric tensor has to be considered a quantum variable. The expected effect is that propagators should be  "smeared out" \cite{ohanian}. In order to preserve the light cone structure, a necessary condition for not violating causality at any instant of the variation of the metric, only conformal fluctuations \cite{padman} are studied here. The averaging over these fluctuations will introduce also a fuzziness of the light cone on the level of the Planck scale.
\vs

According to \cite{mazur} one should not attribute any dynamical meaning to the field describing the conformal deformation of the metric, the apparent kinetic term in the action being interpreted as a potential term expressed in a transformed field variable instead. The fluctuation field does however have a vacuum expectation value, and this is the property used in the present work.
\vs

The fluctuations give rise to a lower bound of the proper length, thus avoiding the infinities mentioned above without the need for regularization. The consequences for several physical quantities related to Green's functions are studied in the present work. The intention here is to check the consistency of the values, with the implicit assumption that the quarks do behave point-like down to the Planck scale (but not beyond that point), and that strong interaction does not introduce an additional (much higher) regularization scale.

\section{Conformal Fluctuations of the Metric}

Let me summarize and adapt in this section some important results of refs. \cite{padman}. The classical quantities and operators are denoted by overbars, e.g. the metric tensor by $\bar{g}_{ij}$. Actually only classically flat space-time shall be considered here, with the signature $\bar{g}_{ij}=\mbox{diag}(1,-1,-1,-1)$. All space-time indices, Latin or Greek, run from 0 to 3. In most expressions, natural units with $\hbar=c=1$ are used.
\vs

The fluctuating metric, subject to a conformal variation, is then written as 
\be
g_{ij}= \bar{g}_{ij}(1+\varphi)^2\equiv\bar{g}_{ij}\Phi^2\ ,
\ee
where the space-time coordinate dependent function $\varphi(x)$ is a scalar field describing the quantum fluctuations of the metric around the classical value. Inserting the full $g_{ij}$ into the (flat space-time) Hilbert action of General Relativity the action becomes
\be
S=-\iv{8\pi^2\lambda^2}\int d^4x \partial^i\phi\partial_i\phi\ ,
\label{S}
\ee
with the reduced Planck length
\be
\lambda=\sqrt{\fr{G\hbar}{3\pi c^3}}\ ,
\ee
$G$ being the gravitational constant. Note that the action above differs from the usual scalar action by a factor 
$-(2\pi\lambda)^{-2}$.
\vs

The vacuum expectation value of the line element becomes
\be
<0|l^2|0>=<0|g_{ij}|0>dx^idx^j=(<(1+\varphi(x))^2>)\bar{g_{ij}}dx^idx^j\ .
\ee
The quantity $<\varphi^2>$ is calculated as the limit $x\ra y$ of the scalar propagator \cite{das}
\be
<\mbox{T}\varphi(x)\varphi(y)>=-(2\pi\lambda)^2 iG(x-y)=-\lambda^2\iv{(x-y)^2-i\epsilon} 
\ \ ,
\ee
with the additional factor mentioned above appropriately included. Then $<\varphi^2>$ diverges as
\be
\lim_{dx\ra 0^+}<\varphi(x+dx)\varphi(x)>=\lim_{dx\ra 0}\fr{\lambda^2}{\bar{g_{ij}}dx^idx^j}\ ,
\label{phi2}
\ee
leading to the result
\be
\lim_{dx\ra 0}<l^2>=\lambda^2\ .
\ee
This means that the fluctuations of the metric impose a lower bound on the proper length. Any point-like object is "smeared out" at the level of the Planck length. Infinities arising from the point-like character of particles are thus avoided.
\vs

Moreover, due to the presence of the fluctuations the classical squared distance $x^2$ is replaced by
\be
<x^2>=x^2+\lambda^2
\label{distance}
\ee
in the expressions considered here. This implies also that after averaging over the fluctuations the light cone is smeared out at the level of the Planck length.

\section{Fermions in Curved Space-Time}

When dealing with fermion propagators one has to take into account that any deviation of the metric from flat space-time alters the Dirac equation. Its general form is \cite{brill,parker}
\be 
(i\gamma^k\nabla_k-m)\Psi=0
\ee
where $\nabla_k$ is the {\sl covariant derivative of a spinor},
\be
\nabla_k\Psi=\partial_k\Psi-\Gamma_k\Psi\ .
\ee
The 4x4 matrices $\Gamma_k$ are obtained from the relation
\be
\partial_k\gamma_i-\Gamma_{ik}^j\gamma_j+\gamma_i\Gamma_k-\Gamma_k\gamma_i=0
\ ,
\label{Gammaeq}
\ee
where the Christoffel symbols
\be
\Gamma_{ki}^j=\iv{2}g^{js}(\dl_k g_{si}+\dl_i g_{sk}-\dl_s g_{ki})
\ee
contain the full metric tensor
\be
g_{ij}=\bar{g}_{ij}\Phi^2=\mbox{diag}(1,-1,-1,-1)\Phi^2\ ,
\ee
and where the $\gamma$ matrices are related to the flat space-time $\bar{\gamma}$ matrices by
\be
\gamma_k=\Phi\bar{\gamma}_k\ .
\ee
A rather lengthy calculation is needed to obtain the explicit form of the matrices $\Gamma_k$ from eq.(\ref{Gammaeq}). Inserting them back into the covariant derivative, the Dirac eq. can be written for the case at hand as
\be
(\slash{p}-m)\Psi=0
\ee
where the operator $\slash p$ reads
\be
\slash{p}=\Phi^{-1}\bar{\gamma}_k \bar{g}^{kk}(\bar{p}_k+i\fr{3}{2}\Phi^{-1}\dl_k\Phi)
\label{slashp}
\ee
with
\be
\bar{p}_k=i\dl_k\ .
\ee

\section{Quark Condensate and Quark Mass}

Let me pass now to the discussion of quantities containing a single fermion propagator. The quark condensate is related to the fermion propagator by
\be
<\bar{q}q>=-i\lim_{y\ra x^+}\iv{N_f}\mbox{Tr}\ g_F(x,y)
\ee
where $N_f$ is the number of flavors and Tr stands for the trace in flavor, color and Dirac spaces. The space-time dependent propagator $g_F(x)$ is obtained from its four-momentum dependent counterpart $G_F(p)$ by a Fourier transform. The condensate becomes
\be
<\bar{q}q>=-3i\lim_{y\ra x^+}\int\fr{d^4\bar{p}}{(2\pi)^4\sqrt{-g}}
e^{-i\bar{p}\cdot(y-x)}\mbox{tr}\iv{\slash{p}-m}\ .
\ee
Here, tr is the Dirac trace  and $m$ the constituent quark mass (see below),  and the $+i\epsilon$ term has been omitted from the denominator, for simplicity of notation. Also implicit is the evaluation of the vacuum expectation value of the fluctuations. The momentum $\slash{p}$ is the full one (\ref{slashp}), and $g$ refers to the determinant of the metric tensor
\be
g=-\mbox{det}\ g_{kl}=\Phi^8\mbox{det}\ \bar{g}_{kl}=-\Phi^8\ .
\ee
To calculate the vacuum expectation value of the fluctuating field, it is convenient to rewrite the denominator in terms of $\bar{p}^2$. After some algebra one has
\be
\mbox{tr}\iv{\slash{p}-m}=\fr{4m\Phi^2(\bar{p}^2-m^2\Phi^2-\fr{3i}{\Phi}\dl_k\Phi\bar{p}_k)}{(\bar{p}^2-m^2\Phi^2)^2}
\ee
and then
\be
<\bar{q}q>=-\lim_{x\ra 0}\fr{12im}{\Phi^2}\int \fr{d^4\bar{p}}{(2\pi)^4}e^{-i\bar{p}\cdot x}\iv{\bar{p}^2-m^2\Phi^2}\ .
\ee
By using the identity
\be
\iv{a+i\epsilon}=-i\int_0^\infty d\alpha e^{i\alpha (a+i\epsilon)}
\ee
(remembering the presence of the $i\epsilon$ in the denominator), the $d^4\bar{p}$-integration can be performed,
\be
<\bar{q}q>=-\lim_{x\ra 0}\fr{3im^2}{2\pi}\int_0^\infty d\alpha
e^{-\fr{im}{2}(\alpha\Phi^2x^2+\iv{\alpha})}=-\fr{3im^2}{2\pi}\int_0^\infty d\alpha
e^{-\fr{im}{2}(\alpha\lambda^2+\iv{\alpha})}\ ,
\ee
where use of (\ref{phi2}) has been made to write the fluctuation average 
\be
\lim_{x\ra 0}<\Phi^2x^2>=\lambda^2\ .
\ee
\vs
The remaining integral evaluates to
\be
<\bar{q}q>=-\fr{3im^2}{\pi^2\lambda}K_1(im\lambda)\ ,
\ee
with $K_1$ being the modified Bessel function.
In the small argument limit, $K_1$ behaves as the inverse of the argument, so one finally gets
\be
<\bar{q}q>=-\fr{3m}{\pi^2\lambda^2}\ .
\label{qbarq}
\ee
\vs

I proceed now to the constituent mass of the quark. An often used effective chiral model lagrangian is due to Nambu and Jona-Lasinio \cite{NJL}. The modern version of the model in Quantum Chromodynamics is reviewed for instance in \cite{klevansky}. The lagrangian of the simplest SU(2) version is written as
\be
L=\bar{\Psi}(i\gamma^\mu\nabla_\mu-m_0)\Psi+k[(\bar{\Psi}\Psi)^2-(\bar{\Psi}\gamma_5\Psi)^2]
\ ,
\ee
where $k$ is the effective strong coupling constant, $m_0$ is the current quark mass and $\Psi$ is the quark field. As one can see, the quark fields are the only degrees of freedom in this model, the information on the gluons is residing in the constant $k$. It has been shown in \cite{thooft} that this type of lagrangian can be obtained from QCD by integrating out the gluonic degrees of freedom. It is reasonable to expect that gluon degrees of freedom are unimportant at "high energy" where we know that quarks are asymptotically free. But how high is "high energy"? Asymptotic freedom is observed in QCD where energy scales are measured in terms of GeV, but since in this work energies up to the Planck energy are considered, there is no guarantee that the present type of lagrangian can still be used, the gluons could become important again. Let me simply {\sl assume}  here the validity of this lagrangian down to the Planck scale and use it to study the consequences of the  fluctuations, remembering the words of caution about its applicability.
\vs

Note that the model is non-renormalizable and is normally defined only together with some regularization procedure, for instance by using a cut-off typically in the order of $1 GeV$, the strong interaction scale. In the present consideration there is no need to regularize, since the results stay finite. The "regularization" arises naturally from the fluctuations on the Planck scale.
\vs

After Fierz symmetriziation in color, flavor and Dirac spaces the lagrangian acquires more terms which are not written explicitely here, since they do not contribute to the quantities to be studied. The coupling constant is redefined by the presence of the exchange terms. Let me call this redefined value $k$.
\vs

Some comments are in order regarding the $\gamma$-matrices appearing in this kind of quartic interaction lagrangian. The expression of the pseudoscalar-isovector term $(\bar{\Psi}\gamma_5\Psi)^2$ stands actually for $(\bar{\Psi}(\gamma_5)^+\Psi)(\bar{\Psi}\gamma_5\Psi)$. Since $\gamma_5=i\gamma^0\gamma^1\gamma^2\gamma^3$ , it carries the inverse fourth power of $\Phi$:
\be
\gamma_5=\Phi^{-4}\bar{\gamma}_5\ .
\ee
The adjoint is however
\be
(\gamma_5)^+=\Phi^{4}\bar{\gamma}_5\ ,
\ee
so the contributions of $\Phi$ cancel in the lagrangian. (The same holds true for terms containing $\gamma_\mu$). It should be noted also that
\be
\bar{\Psi}=\Psi^+\bar{\gamma}^0
\ee
contains only the flat space-time $\bar{\gamma}^0$, in order to be compatible with current conservation 
\be
\nabla_\mu(\bar{\Psi}\gamma^\mu\Psi)=0\ ,
\ee
$\nabla_\mu$ being again the covariant derivative.
\vs

In principle, all the terms appearing in the Fierz symmetric lagrangian have to be considered in the evaluation of the constituent mass. As in the flat space-time case, however, only the scalar term $(\bar{\Psi}\Psi)^2$ contributes, essentially since $\slash{p}$ has the same $\bar{\gamma}_\mu$-structure as $\bar{\slash{p}}$, which cancels the contributions in the traces in the same way, with the exception of the  vector term $(\bar{\Psi}\gamma_\mu\Psi)^2$, see below. The constituent mass $m$ is related to the current mass $m_0$ via the self-energy $\Sigma$:
\be
m=m_0+i\Sigma(p)\ .
\ee
The scalar contribution is
\be
i\Sigma^s=i\lim_{y\ra x}k \mbox{Tr}G_F(y,x)\ .
\ee
This is proportional to the expression for the quark condensate:
\be
i\Sigma^s=-2k<\bar{q}q>\ .
\ee
As mentioned above, the only contribution which does not vanish by tracing is the vector term
\be
i\Sigma^v=i\lim_{y\ra x}k\gamma^\mu\mbox{Tr}G_F(y,x)\gamma_\mu\ .
\ee
After evaluating the trace, this is proportional to $\bar{\gamma}_\mu\int d^4\bar{p}(\bar{p}_\mu\Phi^5+\fr{3}{2}i\Phi^2\dl_\mu\Phi)$, so the first term vanishes right away since it is odd in $\bar{p}_\mu$ in the integral, and the second because it is odd in the quantum fluctuations. Therefore, using eq. (\ref{qbarq}), the constituent quark mass is obtained from
\be
m-m_0=-2k<\bar{q}q>=\fr{6mk}{\pi^2\lambda^2}\ .
\label{mm0}
\ee
This relation can be used to determine the value of the coupling,
\be
k=\fr{\pi^2\lambda^2}{6}\ , 
\label{k}
\ee
with the assumption $m\gg m_0$.

\section{Pion Mass and Weak Decay Constant}

As examples for expressions containing two fermion propagators, consider now the pion in the Nambu--Jona-Lasinio model. The vertex involving the pion is written in terms of quarks as
\be
\fr{g_{\pi q q}}{q^2-m_\pi^2}=k\bar{\gamma}_5\iv{1-kJ(q^2)}\bar{\gamma}_5\ ,
\label{pionvertex}
\ee
with the quark loop integral
\be
J(q^2)=i\int\fr{d^4\bar{p}}{(2\pi)^4\sqrt{-g}}\ \mbox{Tr}\ [\bar{\gamma}_5\tau^-iG_F(p-\iv{2}q)\bar{\gamma}_5\tau^+iG_F(p+\iv{2}q)]\ .
\ee
The integral can be rewritten, with the help of the expression for the constituent quark mass (\ref{mm0}), as
\be
J(q^2)=\iv{k}(1-\fr{m_0}{m})-12iq^2I(q^2)\ ,
\ee
where
\be
I(q^2)=\int\fr{d^4p}{(2\pi)^4}\iv{[(p+\iv{2}q)^2-m^2][(p-\iv{2}q)^2-m^2]}\ .
\ee
Since at the pole $q^2=m_\pi^2$ the quark loop integral obeys $1-kJ(m_\pi^2)=0$, as seen in eq. (\ref{pionvertex}), the pion mass can be deduced from
\be
m_\pi^2=-\fr{m_0}{m}\iv{12ikI(m_\pi^2)}\ .
\label{mpi}
\ee
The integral $I(q^2)$ can be further rewritten using the convolution theorem as
\be
I(q^2)=\int\fr{d^4p}{(2\pi)^4}G_B(p)G_B(p-q)=\int d^4x e^{iq\cdot x}g_B^2(x)\ ,
\label{convolution}
\ee
the functions $G_B(p)$ and $g_B(x)$ being of the form of boson propagators in (four dimensional) momentum and coordinate spaces, respectively. Due to the vacuum expectation value of the fluctuations, $g_B^2(x)$ becomes
\be
g_B^2(x)=\fr{m^2K_1^2(m\sqrt{-(x^2+\lambda^2)}}{(2\pi)^4(x^2+\lambda^2)}.
\ee
Calculating the Fourier transform of $g_B(x)$ and using (\ref{convolution}), one gets
\ba
I(m_\pi^2)=i\fr{\lambda^2}{4\pi^3}
\int_{-\infty}^{\infty}dk_0\int_{0}^{\infty}dk_r
 &\fr{k_r^2K_1(\lambda\sqrt{(k_0-\fr{m_\pi}{2})^2+k_r^2+m^2})K_1(\lambda\sqrt{(k_0+\fr{m_\pi}{2})^2+k_r^2+m^2})}
{\sqrt{(k_0^2+k_r^2+\fr{m_\pi^2}{4})^2+m^4+2m^2(k_0^2+k_r^2-\fr{m_\pi^2}{4})}}\ ,
\ea
having passed to the coordinate system in which $q=(m_\pi,0,0,0)$.
\vs

It is worthwhile noting that, since integrals are finite in the present scheme, it is allowed to shift variables without introducing ambiguities, contrary to the usual case of infinite integrals appearing in normalization or regularization procedures.
\vs

The pion weak decay constant $f_\pi$ in the chiral limit is written as
\be
q_\mu f_\pi^2=-im\int\fr{d^4\bar{p}}{(2\pi)^4
\sqrt{-g}}\mbox{Tr}\ \bar{\gamma_5}\tau^-iG_F(p+\iv{2}q)\iv{2} \gamma_\mu\bar{\gamma_5}\tau^+iG_F(p-\iv{2}q)\ .
\label{qfpi}
\ee
This leads to
\be
f_\pi^2=-12im^2I(0)\ .
\ee
Due to $q^2=0$, the integral $I(0)$ has spherical 4-symmetry and reduces to
\be
I(0)=i\fr{\lambda^2}{(4\pi)^2}\int_{0}^{\infty}dz\fr{z}{z+m^2}K_1^2(\lambda
\sqrt{z+m^2})\ .
\ee
\vs

\section{Numerical Results and Discussion}

Neither the pion weak decay constant nor the quark condensate contain the effective coupling constant $k$ explicitely. In this way, these quantities are model independent, as long as one regards the quarks as the only degrees of freedom. Both quantities are dependent indirectly on $k$ via the constituent quark mass. In order to be able to reproduce the experimental value of the pion weak decay constant $f_\pi=93 MeV$, it is necessary to assume a rather small constituent mass $m=34.6 MeV$. With the more acceptable value $m=386 MeV$, the decay constant becomes $f_\pi=1011 MeV$. The discrepancy may be due to the fact that expression (\ref{qfpi}) is an approximation in the chiral limit. It is interesting to realize that $f_\pi$ is only logarithmically dependent on the regularization point. Using the Planck length $\lambda$ as the covariant cutoff in ref. \cite{blin88} yields very similar results.
\vs

The quark condensate, although finite, evaluates to the unusually large value   $<\nolinebreak\bar{q}q>=(5.5\times10^{15} MeV)^3$, for the choice $m=386 MeV$. As discussed for instance in \cite{blin88}, the quark condensate is not renormalization invariant, but $m_0<\bar{q}q>$ is. In fact, the Gell-Mann--Oakes--Renner relation $-m_0<\bar{q}q>=f_\pi^2m_\pi^2$ indicates that the current quark mass must be very small in the present situation, of the order of $m_0=10^{-39}MeV$, instead of the usual few $MeV$. We can imagine that the "bare" current quark mass, appearing in the lagrangian, is successively dressed by the fluctuations of the metric to the few $MeV$ current quark mass usually encountered in hadron physics, and then by the strong interaction, to finally yield the constituent quark mass $m$. This is seen in eq. (\ref{mm0}), where both the Planck length $\lambda$ and the effective coupling $k$ enter. To be compatible with (\ref{mm0}), the coupling must be very small, $k=10^{-45}MeV^{-2}$ from eq. (\ref{k}). This can also be imagined as the bare value which is then dressed by the fluctuations to give the more usual order of magnitude in the combinations $k(\bar{\Psi}\Psi)^2$ etc. appearing in the lagrangian.
\vs

Turning finally to the pion mass $m_\pi=140MeV$, we find that eq.(\ref{mpi}) is satisfied with the choices $m_0=6.1\times10^{-38}MeV$ for $m=386MeV$, and $m_0=5.7\times10^{-39}MeV$ for $m=34.6MeV$, respectively. Again, the smaller constituent mass seems to be preferred, as it compares well with the Gell-Mann--Oakes--Renner result. One should, however, not attribute too much value to this statement, since the model lagrangian used has to be taken with caution, when approaching the Planck scale. As mentioned before, gluonic degrees of freedom may become increasingly important at very high energies again, since it is not clear if asymptotic freedom holds up to the Planck scale. It may not be allowed to simply use an effective coupling constant to represent strong interactions close to the Planck energy.
\vs

To summarize the results, no major contradictions have been found in the present scheme, in which conformal quantum fluctuations of the metric tensor introduce a fuzziness of point-like particles on the Planck scale. This fact avoids infinities without the need for renormalization or regularization. The physical values of quantities like the pion mass, pion weak decay constant and constituent quark mass attain reasonable values, if one assumes nonstandard (very small) values of the current quark mass and effective strong coupling constant. The fluctuations are then responsible for dressing the naked values of the current quark mass and of the coupling to their more standard values used in low-energy hadron physics.
\vs

\vs

\ni {\bf \Large Acknowledgements}
\vs

\ni I gratefully acknowledge discussions with E. van Beveren, J.L.A. Fernandes, B. Hiller and J. H\"ufner.


\begin{thebibliography}{7}

\bibitem{deser}
S. Deser, Rev. Mod. Phys. {\bf 29} (1957) 417

\bibitem{dewitt}
B.S. DeWitt, Phys. Rev. Lett. {\bf 13} (1964) 114

\bibitem{isham}
C.J. Isham, A. Salam and J. Strathdee, Phys. Rev. {\bf D3} (1971) 867 and 1805, and {\bf D5} (1972) 2548

\bibitem{ohanian}
H.C. Ohanian, Phys. Rev. {\bf D55} (1997) 5140 and {\bf D60} (1999) 104051

\bibitem{padman}
T. Padmanabhan, Phys. Rev. {\bf D28} (1983) 745 and Ann. Phys. (US) {\bf 165} (1985) 38; 

J.V. Narlikar and T. Padmanabhan, {\sl Gravity, Gauge Theories and Quantum Cosmology} (D. Reidel Pub. Co., Dordrecht 1986)

\bibitem{mazur}
P.O. Mazur and E. Mottola, Nucl. Phys. {\bf B341} (1990) 187

\bibitem{das}
A. Das, {\sl Field Theory} (World Scientific, Singapore 1993)

\bibitem{brill}
D.R. Brill and J.A. Wheeler, Rev. Mod. Phys. {\bf 29} (1957) 465

\bibitem{parker}
L. Parker, Phys. Rev. {\bf 183} (1969) 1057 and {\bf D3} (1971) 346

\bibitem{NJL}
Y. Nambu and G. Jona-Lasinio, Phys. Rev. {\bf 122} (1961) 345 and {\bf 124} (1961) 246

\bibitem{klevansky}
S.P. Klevansky, Rev. Mod. Phys. {\bf 64} (1992) 649

\bibitem{thooft}
G. 't Hooft, Phys. Rev. {\bf D14} (1976) 3432

\bibitem{blin88}
A.H. Blin, B. Hiller and M. Schaden, Z. Phys. {\bf A331} (1988) 75

\end{thebibliography}
\end{document}